\documentclass[12pt]{article}
\usepackage{amsmath}
\usepackage{graphicx,psfrag,epsf}
\usepackage{enumerate, color}
\usepackage{natbib, float}
\usepackage{url} 

\usepackage{amsmath,amsfonts}
\usepackage{bm, amssymb}
\usepackage{natbib,theorem}

\newcommand{\red}{\textcolor{black}}
\newcommand{\blue}{\textcolor{black}}
\newcommand{\black}{\textcolor{black}}

\def\v{{\varepsilon}}

\newtheorem{proposition}{Proposition}
\newtheorem{definition}{Definition}

\theoremheaderfont{\hskip\parindent\normalfont}
\theorembodyfont{\normalfont}

\newcommand{\cC}{\mathcal{C}}

\newcommand{\bX}{\mathbf{X}}
\newcommand{\bx}{\mathbf{x}}

\newcommand{\cF}{\mathcal{F}}
\newcommand{\cN}{\mathcal{N}}

\newcommand{\cM}{\mathcal{M}}

\newcommand{\cS}{\mathcal{S}}
\newcommand{\cT}{\mathcal{T}}

\newcommand{\eps}{\varepsilon}

\newcommand{\bbeta}{\boldsymbol{\beta}}

\newcommand{\bgamma}{\boldsymbol{\gamma}}

\newcommand{\E}{\mathrm{E}}

\newcommand{\cov}{\mathrm{cov}}

\newcommand{\sign}{\mathrm{sign}}
\newcommand{\bzero}{\boldsymbol{0}}

\newcommand{\argmin}{\operatornamewithlimits{argmin}}

\newcommand{\blind}{0}

\begin{document}
\def\spacingset#1{\renewcommand{\baselinestretch}%
{#1}\small\normalsize} \spacingset{1}


\if0\blind
{
  \title{\bf A Note on High Dimensional Linear Regression with Interactions}
  \author{Ning Hao and Hao Helen Zhang\thanks{
    The authors gratefully acknowledge the funding support of National Science Foundation DMS-1309507 and DMS-1418172.}\hspace{.2cm}\\
    Department of Mathematics, University of Arizona
    }
  \maketitle
} \fi

\if1\blind
{
  \bigskip
  \bigskip
  \bigskip
  \begin{center}
    {\LARGE\bf A Note on High Dimensional Linear Regression with Interactions}
\end{center}
  \medskip
} \fi

\bigskip

\bigskip
\begin{abstract}
The problem of interaction selection has recently caught much attention in high dimensional data analysis.  This note aims to address and clarify several fundamental issues in interaction selection for linear regression models, especially when the input dimension $p$ is much larger than the sample size $n$. We first discuss issues such as a valid way of defining {\it importance} for the main effects and interaction effects, the invariance principle, and the strong heredity condition. Then we focus on two-stage methods, which are computationally attractive for large $p$ problems but regarded heuristic in the literature. We will revisit the counterexample of Turlach (2004) and provide new insight to justify two-stage methods from a theoretical perspective. In the end, we suggest some new strategies for interaction selection under the marginality principle, which is followed by a numerical example.
\end{abstract}

\noindent%
\vfill

\newpage
\spacingset{1.45} 

\section{Introduction}
\label{sec:Intro}
Given data $\{(\bx_i,y_i)\}_{i=1}^n$, which are independent and identically distributed copies of $(\bX,Y)$, where $\bX=(X_1,...,X_p)^{\top}$ is a $p$-dimensional predictor vector and $Y$ is the response \red{variable}, the \red{standard} linear regression model assumes
\begin{eqnarray}\label{linearmodel}
Y=\beta_0+\beta_1X_1+\cdots+\beta_pX_p+\v.
\end{eqnarray}
In complex systems, the predictors \red{often} work together and their \red{interaction effects can} play a crucial role in \red{model prediction and interpretation}. Historically, models with two- or higher-order \red{interaction terms} have been considered under \red{the standard} linear models and generalized linear models \citep{Nelder:1977, Nelder:1994, MccullaghNelder:1995, McCullagh:2002}, polynomial regression \citep{Peixoto:1987,Peixoto:1990}, experiment designs \citep{HamadaWu:1992,Chipman:1996,ChipmanETAL:1997}, among others. In general, a linear model with two-way interaction \red{effects} is expressed as
\begin{eqnarray}\label{quadraticmodel}
Y=\beta_0+\beta_1X_1+\cdots+\beta_pX_p+\gamma_{11}X_1^2+\gamma_{12}X_1X_2+\cdots+\gamma_{pp}X_p^2+\v,
\end{eqnarray}
where $\beta_0$, $\bbeta=(\beta_1,...,\beta_p)^{\top}$, $\bgamma=(\gamma_{11},\gamma_{12},...,\gamma_{pp})^{\top}$ are unknown parameters. In model \eqref{quadraticmodel}, $X_1$,..., $X_p$ are the \emph{main} effects, $X^2_j$ $(1\leq j\leq p)$ and $X_jX_k$ $(1\leq j\leq k\leq p)$ are \red{the} quadratic and two-way interactions terms, respectively. We refer to all \red{of} the degree-two terms as \emph{interactions} in this note. One \red{special} feature about model \eqref{quadraticmodel} is the intrinsic relationship among the \red{regressor terms}, i.e., $X_jX_{k}$ is \red{a} \emph{child} of $X_j$ and $X_{k}$, and, $X_j$ and $X_{k}$ are \red{the} \emph{parents} of $X_jX_{k}$. This type of model structure is known as {\it hierarchy} or the {\it hierarchical} structure.

In modern biological and medical research, gene-gene interactions, \red{also called} {\it epistatic effects}, and gene-environment interactions \black{have been studied intensively} in genome-wide association studies \red{(GWAS)} \citep{EvansETAL:2006, ManolioCollins:2007, KooperbergLeBlanc:2008,Cordell:2009}. To deal with large and complex data sets, variable selection in regression has been under rapid development over the past two decades; a comprehensive overview is given in \cite{Fan:Lv:Overview:2009} and the book \red{by} \cite{buhlmann2011statistics}. Lately, \red{research on} interaction selection has revived in \red{the context of} high dimensional data analysis; \red{ examples of the recent works} include \cite{LARS:2004}, \cite{ZhaoETAL:2009}, \cite{YuanRoshanZou:2009}, \cite{ChoiLiZhu:2010}, \cite{bien2013lasso}, and \cite{HaoZhang:forward:2012}. For \red{a} large $p$ which is comparable to $n$ or much larger than $n$, \red{the problem of} interaction selection for model (\ref{quadraticmodel}) faces \red{a number of } challenges. Computationally, there are $d=(p^2+3p)/2$ predictors \red{in total}, \red{so} the number of candidate models \red{is} $2^d$, \red{which} can be enormously large and create a bottleneck for standard software. Second, in order to maintain the hierarchical structure
\red{of the final model}, \red{extra} special effort is needed during the selection process. For example, several authors have suggested special penalty functions or constraints to keep the hierarchy \citep{ZhaoETAL:2009, YuanRoshanZou:2009}. However, \red{constrained programming can} become infeasible for large $p$ \red{due to high computational cost}. \red{Theoretically, it is more challenging to study} statistical inferences and asymptotic \red{properties of an estimtor given by an interaction selection method}, since the \red{interaction effects} have more complex covariance structures than \red{the} main effects.

In this note, we first discuss some fundamental issues in interaction selection for model (\ref{quadraticmodel}) \red{in high dimensional settings}. When $p$ is \red{large or extremely large}, two-stage methods might be the only feasible choices in practice. However, there has been a long-term doubt on its theoretical foundation. We aim to \red{shed new light on the validity of} two-stage methods. Finally, we discuss the marginality principle and suggest some new \blue{strategies} suitable for high dimensional interaction selection. Throughout this note, we assume \red{that} $\bX=(X_1,...,X_p)^{\top}$ is a random vector following a continuous distribution \blue{$\cF$}. The noise $\v$ \red{follows} $\cN(0,\sigma^2)$ and \red{it is} independent of $\bX$.

\section{Definition of ``Importance''}
\label{sec:Def}
\red{Consider the question of how to define {\it important effects} in a regression} model. The answer is \red{quite simple} for \red{the standard} linear models containing only the main effects, \red{but not so straightforward for interaction terms} due to \red{the model hierarchy}. In the following, we \red{first review} the invariance principle in \red{the standard} linear models and then suggest a proper definition of {\it importance} for models containing interaction terms.

\subsection{Invariance Principle}

\red{In} model (\ref{linearmodel}), \red{when $p$ is large}, \red{a common model assumption is that} the \red{underlying} true model is sparse, i.e., only a small number of variables are relevant to the response. \red{Naturally, the} {\it relevance} or {\it importance} of a variable $X_j$ is \red{determined by its} coefficient $\beta_j$. Formally, we say that $X_j$ is \emph{important} or \emph{relevant} if and only if $\beta_j\neq 0$.
Variable selection aims to identify all important variables, \red{or in other words}, the support of the coefficient vector $\bbeta$ denoted by $\cS(\bbeta)=\{j: \beta_j\neq 0, j=1, \ldots, p\}$.
For convenience, we define $\sign(\bbeta)=(\sign(\beta_1),...,\sign(\beta_p))^{\top}$.

In real applications, it is a common practice to center or rescale the data before variable selection is conducted. For example, \red{before a shrinkage method like the LASSO \citep{TibshiraniLASSO:1996} is applied}, the predictors are \red{usually} standardized to have zero mean and unit variance, so that they are on the same scale and their regression coefficients are comparable. \red{A proper} definition of ``importance''  should \red{satisfy} the {\it invariance principle} with respect to the coding transformation
of covariates  \citep{Peixoto:1990}. 
To elaborate, consider the transformation $\tilde X_j=a_j(X_j-c_j), j=1,\ldots, p$, where $a_j>0$ and $c_j$
are arbitrary constants. \red{Under this transformation}, model \eqref{linearmodel} becomes
\[Y=\tilde\beta_0+\tilde\beta_1\tilde X_1+\cdots+\tilde\beta_p\tilde X_p+\v=(\beta_0+\sum_{j=1}^p \beta_jc_j) +a_1^{-1}\beta_1\tilde X_1+\cdots+a_p^{-1} \beta_p\tilde X_p+\v .\]
\red{It is clear} that $\tilde\beta_j=a_j^{-1}\beta_j\ne0$ if and only if $\beta_j\ne 0$. \red{Furthermore}, $\sign(\tilde\beta_j)=\sign(\beta_j)$. \red{Therefore}, the definitions of $\cS(\bbeta)$ and $\sign(\bbeta)$ both satisfy the invariance principle.

\red{When studying the theory for high dimensional variable selection}, a \red{number} of model consistency criteria are \red{recently} suggested to study asymptotic properties of a \red{variable selection} procedure, including \blue{sure screening (screening consistency), model selection consistency,} and sign consistency, among others. \red{For a given estimator} $\hat\bbeta$, these three types of consistency amount to, with high probability, $\cS(\hat\bbeta)\supset \cS(\bbeta)$,  $\cS(\hat\bbeta)= \cS(\bbeta)$ and $\sign(\hat\bbeta)=\sign(\bbeta)$, respectively. \red{Due} to the invariance of $\cS(\bbeta)$ and $\sign(\bbeta)$, these consistency properties are also invariant under any coding transformation of the covariates.

\subsection{``Important Effects'' in \red{Models with Interaction Terms}}

We now define the {\it important main effects} and {\it important interaction effects} for model \eqref{quadraticmodel}. This turns out not be so straightforward as in the \red{standard} linear model \eqref{linearmodel}.

First, we point out that, the \red{traditional} definition $\beta_j\neq 0$ \black{or}
$\sign(\beta_j)\neq 0$ for the ``important \red{main effects}'' is no longer proper for model \eqref{quadraticmodel}, since it violates the invariance principle. \red{We illustrate this} by using Turlach's data generating process \citep{Turlach:2004},
\begin{eqnarray}\label{Turlach}
Y=(X_1-0.5)^2+X_2+X_3+X_4+X_5+\v.
\end{eqnarray}
Model \eqref{Turlach} \red{can be expressed in the following three different but} equivalent \red{equations},
\begin{eqnarray*}
Y&=&X_1^2-{\bf 1}X_1+\frac14+X_2+X_3+X_4+X_5+\eps,\\
Y&=&\tilde{X}_1^2+{\bf 0}\tilde{X}_1+X_2+X_3+X_4+X_5+\eps, \qquad\text{with}\quad \tilde{X}_1=X_1-0.5,\\
Y&=&\hat{X}_1^2+{\bf 1}\hat{X}_1+\frac14+X_2+X_3+X_4+X_5+\eps, \qquad \text{with}\quad \hat{X}_1=X_1-1,
\end{eqnarray*}
where the last two \red{expressions are the results of a} simple coding transformation $X_1-c$. In the three expressions, the coefficient of the first main effect is $-1$, $0$ and $1$, respectively. \red{This would} lead to \red{three different} interpretations \red{about} the effect of $X_1$, which is positive, null, or negative. \red{So which one is the correct?} The answer \red{depends on} the coding system. The reason for this inconsistent interpretations is that $X_1^2$ is a function of $X_1$. As long as $\gamma_{jk}\neq 0$, there always exist some transformations to make $\sign(\beta_j)$ or $\sign(\beta_k)$ be positive, negative, or zero.
Furthermore, \red{under \eqref{quadraticmodel}}, neither the support $\cS(\bbeta)$ nor $\sign(\bbeta)$ is invariant of a covariate coding transformation. It is problematic since all of the three expressions correspond to the same model. In general, \red{violating} the invariance principle can be \red{commonly encountered} whenever there is some deterministic intrinsic relationship \red{among the} predictors.

Next, we propose proper definitions for the {\it important effects} in model \eqref{quadraticmodel} \blue{which obey} \red{the invariance principle}.
\begin{definition}
For the data generating process (\ref{quadraticmodel}), we say \red{that} $X_j$ is important if and only if $\beta^2_j+\sum_{k=1}^p\gamma_{jk}^2> 0$, and say $X_jX_k$ is important if $\gamma_{jk}\neq0$. The set of important main effects is defined by $\cT(\bbeta,\bgamma)=\{j: \beta_j^2+\sum_{k=1}^p\gamma^2_{jk}>0, ~ j=1,\ldots, p\}$. The sign of main effects is defined as $\sign(\bbeta)$ under any parametrization with $\E(X_j)=0, j=1,...,p$.
\end{definition}

\red{We} show that Definition 1 is invariant of any coding transformation. Under an arbitrary coding transformation $\tilde X_j=a_j(X_j-c_j)$ with $a_j>0$, we have
 \[Y=(\beta_0+\sum_{j=1}^p\beta_jc_j+\sum_{1\leq j\leq k\leq p}\gamma_{jk}c_j c_k)+\sum_{j=1}^p a_j^{-1}(\beta_j+\sum_{k=1}^p\gamma_{jk}c_k)\tilde X_j +\sum_{1\leq j\leq k\leq p}\gamma_{jk}a_j^{-1}a_k^{-1}\tilde X_j\tilde X_k, \]
where $\gamma_{jk}=\gamma_{kj}$ when $j>k$. \red{Under} the new parametrization, we have
\begin{eqnarray*}
\tilde\beta_0&=&\beta_0+\sum_{j=1}^p\beta_jc_j+\sum_{1\leq j\leq k\leq p}\gamma_{jk}c_j c_k,\\
\tilde\beta_j&=& \sum_{j=1}^p a_j^{-1}(\beta_j+\sum_{k=1}^p\gamma_{jk}c_k),\\
\tilde\gamma_{jk}&=&\gamma_{jk}a_j^{-1}a_k^{-1}.
\end{eqnarray*}
It is easy to \red{check} the following facts:

(i) $\sign(\tilde\gamma_{jk})=\sign(\gamma_{jk})$ 

(ii)
$\beta^2_j+\sum_{k=1}^p\gamma_{jk}^2=0, \Longleftrightarrow  \beta_j=0, \gamma_{jk}=0, ~~ \forall j,k. \Longleftrightarrow  \tilde\beta_j=0,  \tilde\gamma_{jk}=0, ~~ \forall j,k. $

$\hspace{6mm} \Longleftrightarrow  \tilde\beta^2_j+\sum_{k=1}^p\tilde\gamma_{jk}^2=0.$

\vspace{2mm}
\noindent
Throughout this paper, all parameterizations considered are exclusively obtained by a coding transformation from the original data. \red{We can further show that},
the sign of \red{the} main effects is well-defined \red{under Definition 1}. The results are summarized in Proposition 1. \red{Without loss of generality, we treat the following two facts} as equivalent: $X_j$ has a positive sign, or $-X_j$ has a negative sign.

\begin{proposition}
\begin{enumerate}
\item \red{A} main effect $X_j$ is important if and only if $\beta_j\neq 0$ or $\gamma_{jk}\neq 0$ for some $k$, under arbitrary parametrization. \black{In particular, $\cS(\bbeta)\subset\cT(\bbeta,\bgamma)$.}
\item If an interaction effect $X_jX_k, j\neq k$ is important, so are \red{its} parent effects $X_j$ and $X_k$. If $X_j^2$ is important, so is $X_j$.
\item If $\E(X_j)=0$, then under a rescale $\tilde X_j=a_jX_j$ \red{with any} $a_j>0$, $\sign(\tilde\beta_j)=\sign(\beta_j)$.
\end{enumerate}
\end{proposition}

\bigskip
\noindent
\red{The new definitions} of the ``important effects''  \red{given in Definition 1 are valid and well-defined, as they eliminate inconsistent interpretations caused by a coding transformation. More importantly, they provide us a unified framework to} study theoretical properties of a variable selection procedure. In Section 3, we discuss some \red{model selection} consistency results for two-stage methods.

\section{Myths About Two-Stage Methods}
\label{sec:Myths}

Existing procedures for interaction selection can be divided into two categories: {\it one-stage} methods and {\it two-stage} methods. One-stage methods select the main effects and \red{the} interactions simultaneously subject to the hierarchical constraint. \red{They include several recent shrinkage methods which} use asymmetric penalty functions and inequality constraints to keep \red{the} model hierarchy \citep{ZhaoETAL:2009, YuanRoshanZou:2009, ChoiLiZhu:2010,bien2013lasso}.
\red{Their} theoretical properties such as \red{model} selection consistency and the oracle properties have been studied, \red{but mainly} for \red{the} $p<n$ \red{settings. Typically, computational cost of one-stage methods is very high or even} infeasible for large $p$. By contrast, two-stage methods are \red{attractive for high dimensional problems with $p\gg n$ due to their feasible and scalable computation algorithms} \citep{WuTETAL:2009, WuETAL:2010}.
\red{Two-stage methods select the main effects and interaction effects at two separate stages, so their computational cost is much smaller than one-stage methods}.

In practice, two-stage methods are widely used in genomics data analysis. However, \red{they have been
usually} regarded heuristic \red{procedures} in the literature, since their theoretical foundation is not clearly understood. \cite{Turlach:2004} constructed \red{one} counterexample which \red{casts} further doubt on the \red{validity} of two-stage methods. In the following, we re-analyze the counterexample \red{in order} to better understand the mechanism of two-stage methods and why they fail in this case. We then discuss some conditions under which two-stage methods work.

\subsection{Turlach's Counterexample}
A general \red{way of implementing a two-stage method} is as follows: at \red{stage one}, only the main effects are considered for selection; at \red{stage two}, the interaction effects of those main effects which are identified at stage \red{one} are considered for selection. Two-stage methods can retain the hierarchical structure in a natural fashion without involving any complex constraint programming, \red{which explains its computational advantages over one-stage methods}.

\cite{LARS:2004} suggested a two-stage procedure based on \red{the} least angle regression (LARS). At stage one, it selects only \red{the} main effects based on the main-effect model (\ref{linearmodel}). Denote the set of selected main effects by $\widehat\cM\subset\{1,...,p\}$. At stage two, \red{the LARS considers} only the interactions of \red{those} main effects belonging to $\widehat\cM$ \red{and} selects \red{the} interactions based on the following model
\begin{eqnarray}
Y=\beta_0+\sum_{j\in\widehat\cM}\beta_jX_j+\sum_{j,k\in\widehat\cM;~j\leq k}\gamma_{jk}X_jX_k+\v.
\end{eqnarray}
\red{At stage one,} two-stage methods conduct variable selection under a misspecified model (by intentionally leaving out all the interaction effects), which has caused much criticism in the literature on their theoretical justifications. In the discussion of the \red{LARS} paper, \cite{Turlach:2004} constructed a counterexample for which two-stage \red{methods do not} work. The data generating process \red{considered} by \cite{Turlach:2004} is
\begin{eqnarray*}
Y=(X_1-0.5)^2+X_2+X_3+X_4+X_5+\v,
\end{eqnarray*}
\red{which is the same one as} given in equation \eqref{Turlach}. Here $X_1,\ldots, X_{10}$ are \red{independent} and identically distributed from \red{a uniform distribution} Unif$[0,1]$, and they are independent with $\v$. Five variables, $X_1, \ldots, X_5$, are present in model \eqref{Turlach}. Because $\cov(Y,X_1)=0$, the two-stage LARS algorithm of \cite{LARS:2004} can not select $X_1$ at stage \red{one}. \red{Consequently}, the procedure will miss the important quadratic \red{term} $X_1^2$ at stage \red{two.} \blue{In the following two subsections,} \red{we will explain why two-stage methods fail at this example and then discuss under what general conditions two methods would work.}

\subsection{New Insight from Turlach's Example}
\red{For two-stage methods}, based on Definition 1, the key to success is to identify all the important main effects at \red{stage one}, so that all \red{the} important interactions are \red{considered} for selection at stage two. We \red{use} Turlach's example to understand the working mechanism of two-stage \red{methods}.

\red{Without loss of generality}, we first center \red{all the covariates}
$\tilde{X}_j=X_j-\E(X_j)=X_j-0.5, j=1, \ldots, p$ and consider the \red{model in the} following form
\begin{eqnarray}\label{Turlach2}
Y=2+\tilde{X}_1^2+\tilde X_2+\tilde X_3+\tilde X_4+\tilde X_5+\v.
\end{eqnarray}
In \eqref{Turlach2}, the linear term $\tilde X_1$ disappears after the centering transformation. It turns out that no variable selection methods based on \eqref{linearmodel} can identify $X_1$ unless by \red{chance}. To see this with a rigorous analysis, consider the following least squares estimator based on the entire data population,
\begin{eqnarray*}
\bbeta_{LS}&=&\argmin_{\beta_0,...,\beta_5}\E(Y-\beta_0-\sum_{j=1}^5\beta_j\tilde X_j)^2\\
&=&\argmin_{\beta_0,...,\beta_5}\left(\E(2-\beta_0)^2+\E(\tilde{X}_1^2-\beta_1\tilde X_1)^2+\sum_{j=2}^5\E(\tilde X_j-\beta_j\tilde X_j)^2\right)\\
&=&(2,0,1,1,1,1)^{\top}.
\end{eqnarray*}
The second coefficient in $\bbeta_{LS}$ is equal to zero, \red{which} implies that \red{it} is impossible to pick out $X_1$ under model (\ref{linearmodel}), even if we could have observed the entire population.
This explains why two-stage methods \red{fail in selecting $X_1$ in this example}. We \red{point out} that this example is too special to be representative. For example, if we simply change $(X_1-0.5)^2$ to $(X_1-c)^2$ with $c\ne 0.5$ in (\ref{Turlach}), then two-stage methods would be able to identify $X_1$ successfully.

\red{Motivated by Turlach's example, we can establish the general conditions} under which two-stage methods work. Note that the solution produced at stage one by a two-stage method targets on the parameter
\begin{eqnarray}\label{bestlinear}
(\check\beta_0,\check\bbeta)=\argmin_{\beta_0,\bbeta}\E\left(Y-\beta_0-\sum_{j=1}^pX_j\beta_j\right)^2,
\end{eqnarray}
but not on $\bbeta$. Since model \eqref{linearmodel} is misspecified, we \red{do not} expect that $\check\bbeta$ is the same as $\bbeta$ \blue{in general}.
Assume that $\check\bbeta$ is unique and sparse.
Then a necessary condition for two-stage methods to \red{work} is that all \red{the} important main effects $\cT(\bbeta,\bgamma)$ are contained in $\cS(\check\bbeta)$, i.e.
$\cT(\bbeta,\bgamma)\subset\cS(\check\bbeta)$. If a main effect is not in $\cS(\check\bbeta)$, \red{such as} $X_1$ in Turlach's example, then it can not be selected except by \red{chance}.

\red{Is it possible to derive a sufficient} condition \red{which ensures} both $\cS(\bbeta)=\cS(\check\bbeta)$ and $\cT(\bbeta,\bgamma)=\cS(\bbeta)$? If so, it will lead to $\cT(\bbeta,\bgamma)=\cS(\bbeta)=\cS(\check\bbeta)$, \red{which justifies} two-stage \red{methods from a theoretical view point}. Recently, \cite{HaoZhang:forward:2012} \red{gives} a simple and sufficient condition on the data distribution which guarantees $\check\bbeta=\bbeta$. \red{We briefly review the main result here}. Without loss of generality, assume that, in model (\ref{quadraticmodel}), $\E(Y)=0$, $\E(X_j)=0$ for all $j=1, \ldots, p$. Moreover, we \red{also} center \red{all} the interaction \red{terms} and define $Z_{jk}=X_jX_k-\E(X_jX_k)$. Then model  \eqref{quadraticmodel} is equivalent to
\begin{eqnarray}\label{quadraticmodel2}
Y= \beta_1X_1+\cdots+\beta_pX_p+\gamma_{11}Z_{11}+\gamma_{12}Z_{12}+\cdots+\gamma_{pp}Z_{pp}+\v.
\end{eqnarray}
Denote by $\Sigma$ the covariance matrix of vector $(X_1,...,X_p,Z_{11}, ...,Z_{jk},..., Z_{pp})^{\top}$. First, \red{we can show} that, \red{if} the joint distribution of $(X_1,...,X_p)^{\top}$, say, $\mathcal{F}$ is symmetric with respect to the origin $\bzero$, then the covariance matrix $\Sigma$ satisfies
\begin{eqnarray}\label{decomp}
\Sigma=\left(
        \begin{array}{cc}
         \Sigma^{(1)} & 0 \\
          0 & \Sigma^{(2)} \\
          \end{array}
          \right),
\end{eqnarray}
where $\Sigma^{(1)}$ and $\Sigma^{(2)}$ are \red{the} covariance matrices of $(X_1,...,X_p)^{\top}$ and $(Z_{11},...,Z_{pp})^{\top}$, respectively. \red{The block structure is mainly due to the fact} that all \red{the} first and third moments of the joint distribution $\mathcal{F}$ are zero.  The following proposition implies that the block structure of $\Sigma$ \red{is} a sufficient condition for $\check\bbeta=\bbeta$.

\vspace{2mm}

\begin{proposition}\label{p2}
If (\ref{decomp}) holds, then $\check\bbeta=\bbeta$. \black{In particular, $\cS(\bbeta)=\cS(\check{\bbeta})$.}
\end{proposition}

{\bf Proof:} For (\ref{quadraticmodel2}), define $\omega=\gamma_{11}Z_{11}+\gamma_{12}Z_{12}+\cdots+\gamma_{pp}Z_{pp}+\v$. Based on
(\ref{decomp}), we have $\cov(\omega,X_j)=0$ for $1\leq j\leq p$. \blue{Temporally denote by $\bbeta^*$ the true coefficient vector. Then,
\begin{eqnarray*}
\check\bbeta&=&\argmin_{\bbeta} \E\left(Y-\sum_{j=1}^pX_j\beta_j\right)^2\\
&=& \argmin_{\bbeta} \E\left(\sum_{j=1}^pX_j\beta^*_j+\omega-\sum_{j=1}^pX_j\beta_j\right)^2\\
&=& \argmin_{\bbeta} \E\left[\left(\sum_{j=1}^pX_j\beta^*_j-\sum_{j=1}^pX_j\beta_j\right)^2+ \omega^2\right]=\bbeta^*,
\end{eqnarray*}
where the equal sign in the last line holds because all of the variables are centered and $\cov(\omega,X_j)=0$ for all $j$.}
\hfill $\Box$

\vspace{4mm}
\noindent
\red{{\it Remark 1}: The key conclusion from Proposition \ref{p2} is that two-stage methods can identify $\cS(\bbeta)$ successfully at stage one, even if the model is misspecified. The block structure of $\Sigma$ is s sufficient condition for Proposition  \ref{p2}. To satisfy the block structure, one convenient sufficient condition is the symmetry of the joint distribution of $(X_1, \cdots, X_p)^{\top}$; there might be other sufficient conditions. In practice, two-step methods can also handle categorical variables, which are usually recoded into a number of separate and dichotomous variables (the so-called ``dummy coding'').
}

\vspace{3mm}
\noindent
\red{{\it Remark 2}: In high dimensional settings, many predictors tend to be highly correlated with each other. In \cite{HaoZhang:forward:2012}, the performance of two-stage methods are evaluated under various correlation structure settings and the numerical results are promising.}

\vspace{3mm}
\blue{Next, we consider the conditions which guarantee $\cT(\bbeta,\bgamma)=\cS(\bbeta)$.}

\subsection{Strong Heredity Condition}
In literature, heredity conditions were first used in \red{the context of} experiment design \citep{HamadaWu:1992,Chipman:1996,ChipmanETAL:1997}. \red{They have been recently used to study interaction selection in linear regression models} \citep{YuanRoshanZou:2009, ChoiLiZhu:2010}. \red{For} model (\ref{quadraticmodel}) or (\ref{quadraticmodel2}), the strong heredity condition is \red{expressed as}
\begin{eqnarray}\label{heredity}
\gamma_{jk}\ne0\qquad \text{only if} \qquad \beta_j\beta_k\neq0 \qquad \forall \quad 1\leq j,k\leq p.
\end{eqnarray}
\red{And} the weak heredity condition is \red{expressed as}
\begin{eqnarray}\label{wheredity}
\gamma_{jk}\ne0\qquad \text{only if} \qquad \beta_j^2+\beta_k^2\neq0 \qquad \forall \quad 1\leq j,k\leq p.
\end{eqnarray}

\blue{For any fixed} parametrization, the strong heredity condition (\ref{heredity}) implies $\beta_j\ne0$ for any important main effect $X_j$, \red{i.e. $\cT(\bbeta,\bgamma)=\cS(\bbeta)$.} 
\blue{By Proposition \ref{p2}, conditions (\ref{decomp}) and (\ref{heredity}) guarantee that} $\cT(\bbeta,\bgamma)=\cS(\bbeta)=\cS(\check\bbeta)$.


The heredity conditions (\ref{heredity}) and (\ref{wheredity}) seem to be kind of restrictive at the first glance. In the following, we provide some additional insight on the nature of heredity conditions, helping one better understand these conditions.


\red{First, the strong heredity condition is actually not that restrictive, since the set of models which violate the strong heredity condition is usually ``small''. We use a simple setting to illustrate this}. Consider $p=2$ and model (\ref{quadraticmodel2}) \red{with three effects $X_1, X_2, X_1X_2$ (for simplicity, assume no quadratic effects $X_1^2$ and $X_2^2$ are involved)}. The entire parameter space for the coefficient vector $(\beta_1,\beta_2,\gamma_{12})^{\top}$ is $\mathbb{R}^3$. \red{When the strong heredity condition is imposed, it only} excludes \red{the} low dimensional subset $\{\beta_1\beta_2=0,\gamma_{12}^2>0\}$ from $\mathbb{R}^3$. Since this excluded set \red{can be seen as a zero-measure subset of} the Euclidean space, the strong heredity condition \red{essentially} covers the entire model space $\mathbb{R}^3$ almost surely.

\red{Second}, whether a model satisfies heredity conditions does depend on its parametrization. It is a very important fact, which \blue{is however} often overlooked in the literature. \red{In linear regression}, it is a common practice to center or rescale \blue{the data} \red{before fitting the model and conducting variable selection}. \red{Since} any coding transformation $X_j\to a_j(X_j-c_j)$ \red{leads} to a new parametrization for the coefficient vector, \red{it would be} meaningless to discuss heredity conditions of a model without specifying its parametrization. \red{In Turlach's example with a parametrization (\ref{Turlach2}), condition} (\ref{decomp}) holds but (\ref{heredity}) does not. It implies that $\cT(\bbeta,\bgamma)\supsetneqq\cS(\bbeta)=\cS(\check\bbeta)$, \red{which explains why} two-stage \red{methods fail.} 

\red{Third, the \red{definitions} of $\cT(\bbeta,\bgamma)$ and $\cS(\check\bbeta)$ are independent of parametrization.} In other words, the answer to the question whether all \red{the} important main effects are in $\cS(\check\bbeta)$ is irrelevant to the model parametrization. \blue{Nevertheless}, 
a good parametrization helps to 
\blue{to connect these two sets via $\cS(\bbeta)$}.

\blue{In practice, as long as $\cT(\bbeta,\bgamma)=\cS(\check\bbeta)$ holds, two-stage methods can identify all the important main effects at stage one, provided other standard technical conditions. The screening consistency and sign consistency for two-stage methods are recently established by} \cite{HaoZhang:forward:2012} and 
\cite{HaoZhang:lasso:2012}, respectively in the context of forward selection and the LASSO. Similar results should hold for other two-stage methods.

\section{Interaction Selection Under Marginality Principle}
We discussed the theoretical foundation for two-stage methods in the \red{preceding} section. In spite of their validity, two-stage methods have two drawbacks. First, interaction effects \red{are} selected only after the selection of main effects \red{is finished. At stage one}, the noise level is high \red{since} we treat \red{the} interaction effects as noises under a misspecified model. Therefore, it would be difficult to identify weak main effects.
Second, \red{the implementation of} many variable selection procedures \red{requires one to specify a proper tuning parameter} adaptively \red{based on the data}. For two-stage \red{methods}, we need to select the tuning parameter twice, which may cause more errors even if the solution path is correct.
These drawbacks \red{have motivated} us to develop alternative strategies \red{which are feasible} for interaction selection \red{in high dimensional settings.}

\subsection{Marginality Principle}
Historically, \red{the} marginality principle \citep{Nelder:1977} offers an important guidance for variable selection in interaction models. Roughly speaking, the marginality principle requires that any interaction term can be selected only after \red{its} parents enter the model. \cite{Nelder:1994} gave a clear \red{description about the key idea of the principle}.

{\small `` When we fit sequences of quantitative terms such as $x_1$, $x_2$, $x_1x_2$, $x^2_1$, $x^2_2$,..., we have to ask which sequences make sense. if we fit $x_1$ without an intercept, then the response must go through the origin, i.e. zero must be a special point on the $x$-scale where $y$ is zero. Similarly, if $x_1^2$ fitted without an $x_1$ term then the turning-point must occur at the origin (not impossible, but very unlikely). For if $x_1$ might just as well be $x_1-a$ then $(x_1-a)^2=x_1^2-2ax_1+a^2$ and the linear term re-appears. Both terms must be fitted in the order $x_1$, then $x_1^2$, and we say that $x_1$ is $f$-marginal to $x_1^2$. With two continuous variable $x_1$ and $x_2$, new effects arise: if $x_1x_2$ is fitted without $x_1$ and $x_2$ then the response surface must be centered on a col (saddle-point) for the process to make sense. In general there is no reason to expect such a centering to occur, so $x_1$ and $x_2$ must be fitted before $x_1x_2$. ...''}

\red{For} polynomial regression, \cite{Peixoto:1990} argued that a well-formulated model should be invariant under \red{simple} coding transformations. For example, $f(x_1,x_2)=\beta_0+\gamma_{12}x_1x_2$ is not invariant, since one or more linear terms \red{can} show up in the model \red{due to} a coding transformation. The transformation $\tilde{x}_1=x_1-1$ will lead to $f(\tilde x_1,x_2)=\beta_0+\gamma_{12}x_2+\gamma_{12}\tilde x_1x_2$, \red{making it} not sensible to fit \red{the} model $\{1,X_1X_2\}$ without $X_1$ or $X_2$.

Both \red{the} marginality principle and \red{the} invariance principle suggest that the selected model should keep the hierarchical structure. \red{For example, consider model (\ref{quadraticmodel}) with $p=2$}. For simplicity, we tentatively ignore the quadratic terms $X_1^2$ and $X_2^2$. Both the marginality and invariance principles suggest that we should select from the following candidate models: $\{1\}$, $\{1,X_1\}$, $\{1,X_2\}$, $\{1,X_1,X_2\}$, or the full model $\{1,X_1,X_2,X_1X_2\}$; all the other sub-models are not sensible. Note that the marginality principle does not exclude the case that the true data generating process is indeed, say, $Y=1+2X_1X_2+\v$, under \red{a} certain parametrization. In this case, we lose only 2 degrees of freedom to fit the full model. On the other hand, it is risky to fit the model $\{1,X_1X_2\}$ without any priory knowledge. In short, \red{the marginality principle is a good guidance to follow for selecting interaction effects}.

Next, it is worth to point out the difference between \red{the} marginality principle and \red{the} heredity conditions. \red{We regard the former as} a guidance for variable selection in interaction models or other hierarchical models. The selected model \red{must satisfy} the hierarchical structure for any variable selection procedure \red{which employs} the marginality principle. On the other hand, \red{the} heredity conditions put some restrictions on the parameter space, and they depend on the parametrization. They are designed to effectively exclude some undesired data generating processes.

\subsection{Some New Algorithms}
In the aforementioned literatures, there are usually two ways to ensure the hierarchical structure. For one-stage \red{methods}, some carefully designed penalties or inequality constrains \red{are imposed} on $\bbeta$ and $\bgamma$ \red{to} guarantee that the resulted model satisfies the strong heredity condition. For two-stage methods, the hierarchical structure \red{is naturally} preserved \red{due to its selection scheme}. \blue{Here we introduce a new strategy based on the marginality principle.}

Many \red{existing} methods of variable selection produce a family of candidate models which are naturally nested or indexed by a tuning parameter. For example, for a stepwise method such as forward selection and \red{the} LARS, a sequence of nested models is obtained; for a penalization approach \red{such as the LASSO}, a family of models indexed by a tuning parameter is produced. These methods can be directly applied to the standard linear model (\ref{linearmodel}) or the interaction model (\ref{quadraticmodel}) by ignoring the hierarchical structure. \blue{The new strategy utilizes a family of dynamic candidate models $\{\cC_t\}$ lying between models (1) and (2), which initiates at (1) and grows adaptively under the marginality principle. Now we sketch two possible implementations of this strategy. }
For a \blue{forward} selection procedure, we denote by $\widehat\cM_t$ the selected model after step $t$, and set the candidate set $\cC_t$ as all \red{of the} main effects and all \red{of the interaction effects} whose both parents are in $\widehat\cM_t$. \blue{In particular, we set $\widehat\cM_0=\emptyset$ and $\cC_0=\{$all main effects$\}$. At step $t+1$, a forward selection procedure selects one new variable from $\cC_t$ and add it to $\widehat\cM_t$ to obtain $\widehat\cM_{t+1}$.} 
For a penalization procedure \red{like the LASSO}, we denote by $\lambda$ the tuning parameter. \red{The coordinate} decent algorithm \red{is} used to calculate the penalization estimator \blue{along} a discrete sequence $\lambda_{\max}=\lambda_0>\lambda_1>\cdots>\lambda_T>0$. \blue{Again we set  $\widehat\cM_t$ the selected model at step $t$ corresponding to $\lambda_t$ and define $\cC_t$ based on $\widehat\cM_t$ in the same way as above. In the next step with parameter $\lambda_{t+1}$, we conduct coordinate decent algorithm on the candidate model $\cC_t$ to achieve $\widehat\cM_{t+1}$.}
\red{Under this new framework,} we have \red{developed two new methods for interaction selection}; \blue{see \cite{HaoZhang:forward:2012,HaoZhang:lasso:2012}, where the new methods are shown to outperform two-stage methods in numerical studies.}


\section{\red{Numerical Analysis}}

\red{We present a numerical example to illustrate the  performance of two-stage methods for interaction selection in high dimensional linear regression settings. Three methods are considered: two-stage forward selection (two-stage FS), the new forward selection algorithm under the marginality principle (iFORM) described in Section 4.2, and the oracle (Oracle) procedure (which is presented as the gold standard but generally not available in practice). To select the tuning parameter, we use the standard BIC and the extended BIC \citep{chenchen:2008}}. More numerical examples can be found in \cite{HaoZhang:forward:2012}.

\red{Consider a data setting with $n=200$ and $p=1,000$. We generate $\mathbf{X}$ from the multivariate Gaussian distribution with mean $\mathbf{0}$ and the autoregressive correlation $\mbox{Cov}(X_{j}, X_{k})=0.5^{|j-k|}$ for $1\le j,k\le p$. Generate the response $Y$ from model \eqref{quadraticmodel} with $\sigma=2$, the true} \blue{$\bbeta=(2, 0, 2, 0, 2, 0, 2, 0, 2,\mathbf{0}_{991}^{\top})^{\top}$,} \red{$\gamma_{13}=1.5, \gamma_{17}=1.7, \gamma_{57}=1.9, \gamma_{79}=2.1$; the rest of interaction effects are all zero. In this example, the important main effects are $\{X_1,X_3,X_5,X_7,X_9\}$, and the important interaction effects are} \blue{ $\{X_1X_3,X_1X_5,X_5X_7,X_7X_9\}$.}

\red{We run $M=100$ Monte Carlo simulations and report their average performance in selecting the important linear and interaction effects, estimating the nonzero regression coefficients, and making predictions. In particular, to evaluate linear effect selection, we report the probability of identifying the important main effects (Cov), percentage of correct zeros (Cor0), percentage of incorrect zeros (Inc0), and the probability of selecting the set of important main effects exactly (Ext). For interaction selection evaluation, we report the probability of identifying all the important interaction effects (iCov), percentage of correct zeros (iCor0), percentage of incorrect zeros (iInc0), and the probability of selecting the set of important interactions exactly (iExt). We also report the average model size for each method. To evaluate estimation results, we report the mean squared error (MSE) of the estimated regression coefficients and the out-of-sample $R^2$ (Rsq) based on a test set of size $n$ from the same distribution as the data. A larger Rsq suggests a better prediction.}

\red{The numerical results are summarized in Table 1. It shows that that two-stage forward selection method (two-stage FS) works reasonably well in terms of model selection. In particular, it identifies exactly the set of important main effects with 61\% probability and the set of important interaction effects with 48\% probability. This performance is pretty good considering the large dimensionality $p=1000$ and a relatively much smaller sample size $n=200$. The new algorithm iFORM is even better than two-stage FS by identifying exactly the set of important main effects with 96\% probability and the set of important interaction effects with 90\% probability. The size of the final model is 8.19 for two-stage FS and is 9.18 for iFORM, respectively. Note the true model size is 9. 
The MSE is 1.86 for two-stage FS, 0.48 for iFORM, and 0.47 for the oracle method. In summary, the performance of iFORM is very close to that of the oracle procedure.} 

 \begin{table}[H]
 \caption{Numerical results for the simulated example.}
 \begin{tabular}{l|cccc|cccc|ccr}\hline
 &\multicolumn{4}{c|}{Linear Term Selection} & \multicolumn{4}{c|}{Interaction Selection} &\multicolumn{3}{c}{Size and Prediction}\\ \hline
 &Cov&Cor0&Inc0&Ext&iCov&iCor0&iInc0&iExt&size&MSE&Rsq\\ \hline
 two-stage FS &      0.62 &      1.00 &      0.12 &      0.61 &      0.62 &      1.00 &      0.24 &      0.48 &      8.19 &      1.86 &     78.71 \\
 iFORM&      1.00 &      1.00 &      0.00 &      0.96 &      0.99 &      1.00 &      0.00 &      0.90 &      9.18 &      0.48 &     91.30 \\
Oracle&      1.00 &      1.00 &      0.00 &      1.00 &      1.00 &      1.00 &      0.00 &      1.00 &      9.00 &      0.47 &     91.32 \\
 \hline
\end{tabular}
\end{table}

\section{Conclusion}
\label{sec:conc}

This note aims to clarify some important issues \red{in} variable selection for linear model with interactions. The \red{presented} concepts and methods also apply to generalized linear models and models with higher-order interaction terms \red{or} complex hierarchical structures. \red{In practice, when} choosing between main effect models, two-way interaction models, or higher-order interaction models, one needs to consider the bias-variance tradeoff. In general, adding \red{more} interaction terms to the model \red{tends to reduce the modeling} bias but increase the variance. 

\bibliographystyle{biometrika}
\bibliography{interaction}
\end{document}